\documentclass{elsart1p}

\usepackage{graphics,amssymb}

\usepackage{latexsym}
\usepackage{amsbsy}
\usepackage{amsfonts}
\usepackage{amssymb}
\usepackage{cite}
\usepackage{color}
\usepackage{epsfig}

\mathchardef\Gamma="0100
\mathchardef\Theta="0102
\mathchardef\Lambda="0103
\mathchardef\Xi="0104
\mathchardef\Pi="0105
\mathchardef\Sigma="0106
\mathchardef\Upsilon="0107
\mathchardef\Phi="0108
\mathchardef\Psi="0109
\mathchardef\Omega="010A

\def\I{{\rm i}}

 \def\cal{\mathcal}
 
 \def\vec{\boldsymbol}
\def\q{\rm} 

\def\cha{\kern .6em{\sqcup \kern -1.16em \sqcup}\kern .6em}

\begin{document}
\begin{frontmatter}
\null \vskip -3cm
\title{\bf Quaternionic representation and design \\ of  depolarizers}
\author{Pierre Pellat-Finet}
\address{Universit\'e Bretagne Sud,  UMR  CNRS  6205, LMBA, F-56000 Vannes, France \\ pierre.pellat-finet@univ-ubs.fr}


\begin{abstract}
Quaternions have been used to represent polarization states
  and polarization operators. But so far, only polarizers, dichroic or non-depolarizing
  devices have been represented in that way. We propose a quaternionic representation
  of perfect as well as partial depolarizers. It leads us to  design actual setups for producing
  natural (unpolarized) light from an arbitrary partially or completely polarized wave or for reducing the degree of polarization of a wave. We
  conclude by making the link with Azzam's polarization orthogonalization problem.
 

\end{abstract}

\end{frontmatter}


\section{Introduction}

Quaternions have been used to represent polarization states as well as polarization ope\-rators (birefringent and dichroic devices, polarizers) \cite{PPF1,PPF2,PPF3}. Their introduction in polarization optics results from essential properties of polarized light; in particular, a Lorentz quadratic form (Lorentz metric), built on appropriate magnitudes, plays a preponde\-rant role and leads us to consider  polarization states as endowed with the structure of Minkowski's space. Polarization operators are then described as Lorentz proper rotations whose group is classically related to quaternions \cite{PPF3,PPF4,PPF5}.
A simpler  and more direct introduction, less profound, can be developed from the Stokes vector theory \cite{PPF1,PPF2}, so that the considered quaternionic representation of polarization optics can be seen as an alternative to the Stokes-Mueller formalism. It can also be regarded like an analytic tool complementing the Poincar\'e sphere.

In the present article we extend the above mentioned formalism to depolarizers. The quaternionic representation of a depolarizer will lead to an actual design of such a device.

\section{Quaternionic representation of polarized light}
\subsection{Stokes vectors and Poincar\'e sphere}
For a short and simple introduction to the quaternionic representation of polarization states, we refer to Stokes vectors and to the Poincar\'e sphere \cite{Bor,Ram}. A Stokes vector is written $\vec X=\,^t(X_0,X_1,X_2,X_3)$, where subscript $t$ means ``transpose'' and where the $X_\mu$'s, 
called Stokes parameters,  are positive quantities, homogeneous to irradiances (up to a mutiplicative dimensional factor, equal to the impedance of the propagation medium) or ``vibratory intensities'' \cite{PPF0}. Stokes parameters generally refer to a plane wave with a given direction of propagation; polarization states may be regarded as complex vectors in a plane wave-surface. A Stokes vector $\vec X$  associated with a real (physical) lightwave ${\cal X}$ is such that $(X_0)^2-(X_1)^2-(X_2)^2-(X_3)^2\ge 0$. The lightwave is completely polarized if $(X_0)^2=(X_1)^2+(X_2)^2+(X_3)^2$; it is unpolarized (natural light) if $X_1=X_2=X_3=0$; and  partially polarized if  $(X_0)^2>(X_1)^2+(X_2)^2+(X_3)^2$. The degree of polarization of the lightwave is $\rho$  ($0\le \rho\le 1$), such that ($X_0\ne 0$)
\begin{equation}
  \rho={\sqrt{(X_1)^2+(X_2)^2+(X_3)^2}\over X_0}\,.\end{equation}

In the following we write ``polarization state'' of a lightwave, in the general meaning of partial-polarization state, including completely polarized states, also called pure states, and unpolarized states, as particular cases. Since Stokes parameters measured on a lightwave are proportional to the wave power, by abuse we also call ``power'' the parameter $X_0$, when we refer to a lightwave.

If $\vec X=\,^t(X_0,X_1,X_2,X_3)$ is the Stokes vector associated with a completely polarized lightwave, then
$\vec X_{\! \perp}=\,^t(X_0,-X_1,-X_2,-X_3)$ is associated with an orthogonally (and completely) polarized lightwave.

Stokes vectors can be added \cite{Ram}: if $\vec X$ and $\vec Y$ respectively represent the polarization states of lightwaves ${\cal X}$ and ${\cal Y}$, then $\vec X+\vec Y$ represents the polarization state of the incoherent superposition of waves ${\cal X}$ and ${\cal Y}$, symbolically written ${\cal X}+{\cal Y}$.

\begin{figure}[b]
\begin{center}
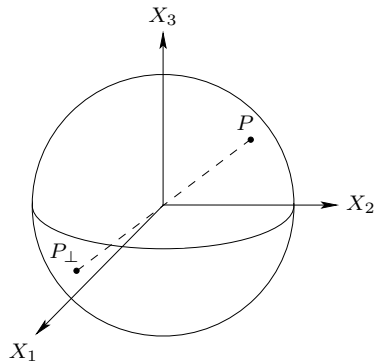
\end{center}
\caption{The Poincar\'e sphere. Points $P$ and $P_\perp$, opposite on the sphere, represent two orthogonally polarized lightwaves.\label{fig1}}
\end{figure}

Pure states are represented on the so-called Poincar\'e sphere \cite{Bor,Ram}. For a pure state, since  $(X_0)^2=(X_1)^2+(X_2)^2+(X_3)^2$, the knowledge of $X_1$, $X_2$ and $X_3$ completely characterizes the polarization state.   The sphere is referred to 3 orthogonal axes, each axis corresponding to a Stokes parameter $X_j$ ($j=1,2,3)$ (Fig.~\ref{fig1}).
A pure polarization state is represented by a point $P$ on the sphere, and since the radius of the Poincar\'e sphere is generally taken equal to 1, that is $X_0=1$,  Stokes parameters $X_1$, $X_2$ and $X_3$ are the Cartesian coordinates of $P$ on the sphere.

The intersection of the sphere with  the plane $X_1$--$X_2$ is called equator. Points on the equator correspond to rectilinear polarization states. 
In general, the point $P (1,0,0)$ represents the ``horizontal'' polarization (in the wave surface), and point $P'(-1,0,0)$ the ``vertical'' one. (Lightwave direction of propagation is also assumed to be horizontal; it is  orthogonal to both  horizontal and vertical polarization directions.) The North pole $(0,0,1)$ represents a left-handed circular polarization, and the South pole $(0,0,-1)$ a right-handed circular polarization.  If $P$ represents the polarization of the wave ${\cal P}$, the point $P_\perp$, opposite to $P$ on the sphere, represents the wave ${\cal P}_\perp$ that is orthogonally polarized to ${\cal P}$ (with the same power).

The Poincar\'e sphere is helpful  in managing birefringent media, since the effect of a birefringent device on  polarization states is a rotation on the sphere. The angle of the rotation is the birefringence (that is, the phase shift between eigenvibrations) and the axis is $E_\perp E$, where $E$ represents the  fast eigenvibration, on the sphere, and $E_\perp$ the slow eigenvibration  of the device ($E_\perp E$ constitutes a diameter of the sphere).

\subsection{Quaternions}

We now consider the 4--dimensional complex vector space ${\mathbb C}^4\!$, whose canonical basis is
\begin{equation}
  {\q e}_0=(1,0,0,0)\,,\;\;\;
  {\q e}_1=(0,1,0,0)\,,\;\;\;
  {\q e}_2=(0,0,1,0)\,,\;\;\;
  {\q e}_3=(0,0,0,1)\,,\end{equation}
and we define a quaternionic product by
\begin{eqnarray}
  &&\!({\q e}_0)^2=-({\q e}_1)^2=-({\q e}_2)^2 =-({\q e}_3)^2={\q e}_0\,, \;\;\;  \\
  &&{\q e}_0\,{\q e}_j={\q e}_j={\q e}_j\,{\q e}_0\,,\;\; j=1,2,3\,, \\
 &&{\q e}_1\,{\q e}_2={\q e}_3=-{\q e}_2\,{\q e}_1\,,\;\,\mbox{and circular permutations  on $j= 1,2,3$.}
\end{eqnarray}
The quaternionic product is not commutative; it is distributive with respect to addition. Equipped with the quaternionic product,  ${\mathbb C}^4$ becomes the algebra of complex quaternions, denoted ${\mathbb H}_{\rm c}$.

A complex quaternion is written $q=q_0{\q e}_0+q_1{\q e}_1+q_2{\q e}_2+q_3{\q e}_3$, where the $q_\mu$'s 
are complex numbers.

Since ${\mathbb H}_{\rm c}$ can be identified with ${\mathbb C}\oplus{\mathbb C}^3$, we have ${\q e_0}\equiv 1$, so that  we often write $1$ in place of ${\q e}_0$.

\subsubsection*{Some noteworthy quaternions}

\smallskip
The Hamilton conjugate of  $q=q_0{\q e}_0+q_1{\q e}_1+q_2{\q e}_2+q_3{\q e}_3$ is
$q^*=q_0{\q e}_0-q_1{\q e}_1-q_2{\q e}_2-q_3{\q e}_3$, and its complex conjugate is
$\overline{q}=\overline{q_0}{\q e}_0+\overline{q_1}{\q e}_1+\overline{q_2}{\q e}_2+\overline{q_3}{\q e}_3$. Then $\overline{q^*}={\overline{q}}^*$.

The norm of $q$ is $N(q)=q\,q^* =q^*\,q=(q_0)^2+(q_1)^2+(q_2)^2+(q_3)^2$. The norm of quaternions is multiplicative: if $q$ and $r$ are quaternions, then $N(qr)=N(q)\, N(r)$.

A quaternion is unitary (or a unit quaternion) if $N(q)=1$.
A quaternion $q$ is a pure quaternion if $q_0=0$. For example the ${\q e}_j$'s, $j=1,2,3$, are unit pure quaternions.
A quaternion
${\q e}_n=n_1{\q e}_1+n_2{\q e}_2+n_3{\q e}_3$, with $(n_1)^2+(n_2)^2+(n_3)^3=1$, is a unit pure quaternion; it is such that
$({\q e}_n)^2=-{\q e}_0$.

The exponential of the quaternion $q$ is
\begin{equation}
  \exp q=\sum_{j=0}^{+\infty}{q^j\over j!}\,.
\end{equation}
If $q\;=\;\psi \,{\q e}_n$, where $\psi$ is a complex number and ${\q e}_n$ a unit pure quaternion, we have
$({\q e}_n)^2=-{\q e}_0$, and we obtain
  \begin{equation}
    \exp {\q e}_n\psi={\q e}_0\sum_{j=0}^{+\infty}(-1)^j{\psi^{2j}\over (2j)!}+{\q e}_n\sum_{j=0}^{+\infty}(-1)^j{\psi^{2j+1}\over (2j+1)!}
    ={\q e}_0\cos\psi +{\q e}_n\sin\psi\,.
    \end{equation}

  \subsubsection*{Matrix representation}

  \smallskip

  Complex quaternions can be represented by $2\times 2$ matrices, with the help of Pauli matrices
  \begin{equation}
  \sigma_0=\begin{pmatrix}{ 1 & 0\cr 0 & 1} \end{pmatrix}\,,\;\;\;
  \sigma_1=\begin{pmatrix}{ 1 & 0\cr 0 & -1} \end{pmatrix}\,,\;\;\;
  \sigma_2=\begin{pmatrix}{ 0 & 1\cr 1 & 0} \end{pmatrix}\,,\;\;\;
  \sigma_3=\begin{pmatrix}{ 0 & -\I\cr \I & 0}\,.
 \end{pmatrix}
  \end{equation}
The correspondence is
\begin{equation}
  {\q e}_0\longmapsto \sigma_0\,,\;\; {\q e}_j\longmapsto -\I\sigma_j\,,\;\;j=1,2,3\,,\end{equation}
so that
\begin{equation}
  q=q_0{\q e}_0+q_1{\q e}_1+q_2{\q e}_2+q_3{\q e}_3\longmapsto
  \sigma=q_0{\q e}_0-\I q_1\sigma_1-\I q_2\sigma_2-\I q_3\sigma_3\,.\end{equation}
We have
\begin{equation}
  \sigma =\begin{pmatrix}{q_0- \I q_1 & \;& -q_3-\I q_2\cr \cr
  q_3-\I q_2  & &q_0+ \I q_1
  }\end{pmatrix}\,.
  \end{equation}
If ${\rm det}\,\sigma $ denotes the determinant of matrix $\sigma$, then $N(q)=\mbox{det}\,\sigma $.

\subsection{Representing polarization states and operators by quaternions}

\subsubsection*{Polarization states}

\smallskip

We represent the Stokes vector $\vec X=\,^t(X_0,X_1,X_2,X_3)$ by the quaternion
\begin{equation}
  X=X_0{\q e}_0+\I (X_1{\q e}_1+ X_2{\q e}_2+ X_3{\q e}_3)\,.\end{equation}
According to Synge, such a quaternion is called a minquat \cite{Syn,PPF3}.
It  can also be written
\begin{equation}
  X=X_0({\q e}_0+\I\rho \,{\q e}_n)=X_0(1+\I\rho \,{\q e}_n)\,,\end{equation}
where $\rho$ is the degree of polarization of the corresponding polarization state (if $X_0\ne 0$) and where
\begin{equation}
  {\q e}_n={1\over \rho X_0}(X_1{\q e}_1+X_2\,{\q e}_2+X_3\,{\q e}_3)\,,
\end{equation}
if $\rho\ne 0$. (If $X_1=X_2=X_3=0$,  then $\rho=0$ and ${\q e}_n$ is not determined.)

\subsubsection*{Polarized and unpolarized components}

\smallskip

With abuse, we also called ``polarization state'' a minquat such as $X_0(1+\I\rho \,{\q e}_n)$, which represents a lightwave polarization-state.

The minquat $X=X_0(1+\I\, {\q e}_n)$ is a completely polarized state to which  $X_{\!\perp}=X_0(1-\I {\q e}_n)$  is orthogonally polarized.

The (completely) polarized component of $X=X_0(1+\I\rho\,{\q e}_n)$ is $X_{\rm c}=\rho X_0(1+\I \,{\q e}_n)$, which owns a power $\rho X_0$; the  unpolarized component of state $X$ is $X_{\rm u}=(1-\rho)X_0$. If $X'=X_0(1-\I\rho \,{\q e}_n)$, then $X'_{\rm c}=X_{{\rm c}\perp}$ and $X'_{\rm u}=X_{\rm u}$: the polarized components of $X$ and $X'$ are orthogonal to each other, whereas their unpolarized components  are equal.

\subsubsection*{Birefringent and dichroic devices}

\smallskip
A birefringent device---generally an elliptical one---has two (unitary) eigenvibrations, represented by two opposite points on the Poincar\'e sphere. Those two points form a diameter of the sphere, called the birefringent axis. The phase shift between the slow and fast eigenvibrations, say $\varphi$, is called the birefringence. On the Poincar\'e sphere, the effect of a birefringent device is a rotation of angle $\varphi$ around the birefringent axis (oriented from the slow vibration representative point  towards the fast vibration representative one).

In the following, we identify a pure unit quaternion ${\q e}_n=n_1{\q e}_1+n_2{\q  e}_2+n_3{\q e}_3$ with the vector joining the center $O$ of the sphere to the point of coordinates $(n_1,n_2,n_3)$.  The birefringent axis (as previously defined)  is then perfectly determined by a unit pure quaternion, say ${\q e}_n$, also called by abuse the axis of the birefringent. 

Finally, the birefringent device of axis ${\q e}_n$ and birefringence $\varphi$ is represented by the unit quaternion
$\exp ({\q e}_n\varphi /2)$. For an incident polarization state represented by the minquat $X$, the
outgoing state is represented by $X'$ such that \cite{PPF1,PPF2,PPF3}
\begin{equation}
  X'=\exp \left({\q e}_n{\varphi \over 2}\right)\,X\,\exp \left(-{\q e}_n{\varphi \over 2}\right)\,.\label{eq11}\end{equation}
(Equation (\ref{eq11}) is shown to express a rotation operating on minquats.)

A dichroic device whose axis is ${\q e}_n$, dichroism $\delta$ and
isotropic absorption $\alpha$ ($\alpha >0$), is represented by
$\sqrt{\alpha} \exp (\I {\q e}_n\delta / 2)$  and its effect on $X$ is expressed by
\begin{equation}
  X'=\alpha \exp \left(\I {\q e}_n{\delta \over 2}\right)\,X\,\exp \left(\I {\q e}_n{\delta \over 2}\right)\,.\end{equation}
If $\alpha =1$, the device is a pure dichroic device. Then $X'_0>X_0$, which means that an eigenstate is amplified, whereas the orthogonal state is attenuated. Active cavities of some lasers may be regarded as pure dichroic devices.
(Choosing the word ``absorption'' for $\alpha$ does not sound  very appropriate!)

\subsubsection*{Polarizers}

\smallskip
We complete the quaternionic representation of polarization optics by describing polarizers. A pure (or ideal) polarizer lets a specific completely polarized wave  pass and stops the orthogonally polarized waves. If the passing state  corresponds to the minquat $X_0(1+\I e_n)$, the polarizer is represented by $(1+\I e_n)/2$.
If  the incident state is $X$, 
the outgoing state is
\begin{equation}
  X'={1\over 4}\,(1+\I e_n)\,X\,(1+\I e_n)\,.\end{equation}
We also say that ${\q e}_n$ is the axis of the polarizer (on the Poincar\'e sphere).

For example, if $X=X_0(1+\I\,{\q e}_n)$, we obtain $X'=X$; if  $X=X_0(1-\I\,{\q e}_n)$, then $X'=0$.

If  $X=X_0(1+\I\rho\,{\q e}_m)$ and if $\theta$ is the angle between ${\q e}_n$ and ${\q e}_m$ (on the Poincar\'e sphere), then
\begin{equation}
  X'={X_0\over 2}(1+\rho \cos\theta )(1+\I\,{\q e}_n)\,.\end{equation}
The outgoing wave is completely  polarized along the passing state of the polarizer (i.e. along the axis ${\q e}_n$ on the Poincar\'e sphere), and its power is $X_0(1+\rho\cos\theta )/2$, which is a generalized form of Malus law.

\section{Representation of depolarizers}

\subsection{Complete depolarizers}

The  state outgoing from a polarizer is completely polarized (or it is 0), whatever the
incident state maybe. At first, a complete (or perfect) depolarizer device could be thought of as the inverse
operator of a pure polarizer: whatever the input state is, the output should
be a completely-unpolarized wave (natural light).
But a polarizer is not an 
invertible operator and the previous representation of a polarizer cannot be used 
to represent a depolarizer (a quaternion of the form $1+\I {\q e}_n$ has no inverse in ${\mathbb H}_{\rm c}$).

On the other hand, we notice that a dichroic device is a partial polarizer. If we send  an
 unpolarized wave, represented by $X=X_0$, on such a device, the output
state corresponds to a partially polarized wave whose  polarized component  is  along the dichroic axis, say ${\q e}_n$ (on the Poincar\'e sphere). The quaternion
$\sqrt{\alpha} \exp (\I {\q e }_n\delta /2)$ is invertible: its inverse is
$(1/\sqrt{\alpha} ) \exp (-\I {\q e}_n\delta /2)$. But the corresponding  operator does not conform to the notion
  of partial depolarizer, because it has not identical effects on two arbitrary
  states whose degrees of polarization are equal, as a perfect depolarizer
  should have. Indeed,  the inverse operator
  of a dichroic device is another dichroic device, multiplied by an isotropic factor.

A perfect depolarizer is  such that for every incident wave the outgoing
wave is unpolarized. For every  input minquat $X$,  the output minquat should take the form  $X'=X'_0$.

Let us consider the (linear) mapping $\psi$ defined for every quaternion $q$ by
\begin{equation}
\psi (q)={1\over 4}(q-{\q e}_1\,q\,{\q e}_1-{\q e}_2\,q\,{\q e}_2-{\q e}_3\,q\,{\q e}_3)\,.\end{equation}
We have $\psi (1)=1$ and
$\psi (e_1)=\psi (e_2)=\psi (e_3)=0$,
and, by linearity, for every minquat $X=X_0(1+\I \rho {\q e}_n)$,  we obtain 
$\psi (X)=X_0$,
and that is exactely the effect of a perfect depolarizer. In conclusion, the mapping, defined on minquats by
\begin{equation}
  D\;:\;X\longmapsto D(X)={1\over 4}(X-{\q e}_1\,X\,{\q e}_1-{\q e}_2\,X\,{\q e}_2-{\q e}_3\,X\,{\q e}_3)\,,\label{eq19}\end{equation}
represents a perfect depolarizer.

More generally, let $[e_u, e_v,e_w]$ be an orthonormal basis of the subspace of
pure quaternions. Then the mapping defined for every minquat $X$ by
\begin{equation}
 D'\;:\;X\longmapsto  D'(X)={1\over 4}(X-{\q e}_u\,X\,{\q e}_u-{\q e}_v\,X\,{\q e}_v-{\q e}_w\,X\,{\q e}_w)\,,\label{eq20}\end{equation}
represents a perfect depolarizer.

\bigskip
\noindent {\em Remark 1.} If $X=X_0(1+\I\rho\, {\q e}_n)$, then $-{\q e}_1X{\q e}_1-{\q e}_2X{\q e}_2-{\q e}_3X{\q e}_3=X_0(3-\I\rho\,{\q e}_n)$, whose polarized component $\rho X_0(1-\I\,{\q e}_n)$ is orthogonal to $\rho X_0(1+\I\,{\q e}_n)$, the polarized component of $X$. According to Eq.\ (\ref{eq19}), the depolarizing effect is obtained by adding two lightwaves with orthogonal polarized components. (The same conclusion holds with $[{\q e}_u,{\q e}_v, {\q e}_w]$ in place of $[{\q e}_1,{\q e}_2,{\q e}_3]$.)

\subsection{Partial depolarizers}

A partial depolarizer is a device that reduces the degree of polarization of every incident wave by a given factor. More precisely, let $\gamma$ be a real number with $0\le \gamma\le 1$.  
A partial depolarizer  depolarizes by a factor $1-\gamma$, if the output
corresponding to the input minquat $X=X_0(1+\I\rho\, {\q e}_n)$, is $X_0(1+\I\gamma \rho \, {\q e}_n)$; the degree of polarization $\rho$ becomes $\gamma \rho$. (A perfect depolarizer depolarizes by a factor 1, that is, $1-\gamma =1$, and $\gamma =0$: for every incident wave the outgoing wave is unpolarized.)

In Eq.\ (\ref{eq20})---and similarly in Eq.\ (\ref{eq19})---the polarized component of $X$ is balanced by the polarized component of $-{\q e}_uX{\q e}_u-{\q e}_vX{\q e}_v-{\q e}_wX{\q e}_w$. To obtain a partially polarized lightwave, not necessarily an unpolarized one, that is, a wave whose polarized component is not necessarily zero, we reduce the weight of $-{\q e}_uX{\q e}_u-{\q e}_vX{\q e}_v-{\q e}_wX{\q e}_w$ by introducing a factor $\beta$ ($0\le \beta\le 1$), and 
we consider the mapping
\begin{equation}
D_{\rm p}\; :\; X\longmapsto D_{\rm p} (X)={1\over 1+3\beta}\,\bigl[X-\beta({\q e}_u\,X\,{\q e}_u+{\q e}_v\,X\,{\q e}_v+{\q e}_w\,X\,{\q e}_w)\bigr]\,.\label{eq22}\end{equation}
Then
\begin{equation}
D_{\rm p} \bigl[X_0(1+\I \rho \,{\q e}_n)\bigr]=X_0\left(1+{1-\beta\over 1+3\beta}\I\rho\,  {\q e}_n\right)\,.\end{equation}
The degree of polarization of $D_{\rm p}[X_0 (1+\I \rho\, {\q e}_n)]$ is
$\rho (1-\beta )/(1+3\beta )$. Thus, to reduce the degree of polarization from
$\rho$ to $\gamma\rho$, we choose $\beta$ such that
\begin{equation}
{1-\beta \over 1+3\beta}=\gamma\,,\end{equation}
and we obtain
\begin{equation}
\beta ={1-\gamma \over 1+3\gamma}\,.\end{equation}
We report the value of $\beta$ in Eq.\ (\ref{eq22})  and eventually write
\begin{equation}
D_{\rm p} (X)={1+3\gamma \over 4}X-{1-\gamma\over 4}({\q e}_u\,X\,{\q e}_u+{\q e}_v\,X\,{\q e}_v+{\q e}_w\,X\,{\q e}_w)\,,\end{equation}
for the general representation of a (generally partial) depolarizer whose depolarizing factor is
$1-\gamma$.

\section{Designing depolarizers}

\subsection{Perfect depolarizers}

\begin{figure}[b]
\begin{center}
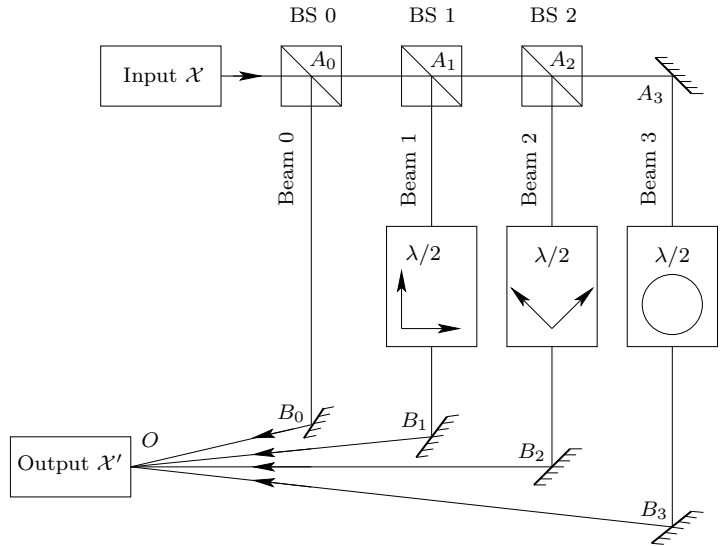
\caption{Setup for a perfect depolarizer. For each half-wave plate the eigenvibrations are indicated.\label{fig2}}
\end{center}
\end{figure}

For the sake of simplicity, we consider $[e_u, e_v,e_w]=[e_1, e_2,e_3]$, and we begin by
the design of a perfect depolarizer ($\gamma =0$). The mapping
$X\longmapsto X$ is the identity. Since $\exp ({\q e}_1\pi /2)={\q e}_1$, according to Eq.\ (\ref{eq11}) the mapping $X\longmapsto -{\q e}_1\,X\,{\q e}_1$
corresponds to the effect of a half-wave plate whose axis is the $X_1$--axis
(on the Poincar\'e sphere), or ${\q e}_1$ with quaternions. According to Eq.\ (\ref{eq19}), we obtain a perfect depolarizer device by sending
the input polarized wave on three half-wave plates (with respective
axes along ${\q e}_1$, ${\q e}_2$ and ${\q e}_3$) and  by making the incoherent
sum of the three corresponding outgoing waves, plus the incident wave. The process is
schematized on Fig.\ \ref{fig2}.

The half-wave plate whose axis is ${\q e}_1$ ($X_1$) on the
Poincar\'e sphere has its fast and slow
 vibrationss horizontal and vertical. They are at $\pm 45^{\circ}$ for the half-wave plate whose axis is along ${\q e}_2$. The third half-wave plate is optically active:
its eigenvibrations are circularly polarized as schematically indicated on Fig.~\ref{fig2} (the rotarory power is half the birefringence, that is, $\pi /2$). Beam splitter BS 0 has a transmission factor $T_0=0.75$, whereas its reflection factor is $0.25$. For BS 1, we have $T_1=0.67$ and $R_1=0.33$; for  BS 2, we have $T_2=0.50$ and $R_1=0.50$. Beams 0, 1, 2, 3 thus have identical powers ($0.25$ times the input power). For the wave superposition at $O$ to be incoherent, the optical path differences $[A_2A_3B_3O]-[A_2B_2O]$, $[A_1A_2B_2O]-[A_1B_1O]$ and $[A_0A_1B_1O]-[A_0B_0O]$ must be greater than the coherence length of the light source.

\subsection{Partial depolarizers}

We assume $[{\q e}_u,{\q e}_v,{\q e}_w]=[{\q e}_1,{\q e}_2,{\q e}_3]$. For the design of a partial depolairzer, we  just adapt beam splitter BS 0. Its transmission factor becomes $T_0= 3(1-\gamma)/4$ and its reflection factor $R_0=(1+3\gamma )/4$.

\bigskip
\noindent{\em Remark 2.} A more general partial polarizer would a priori reduce the degree of polarization,  but also change the polarization of the polarized component of the incident wave.  Such a device actually involves a partial depolarizer in the previous meaning (that is, without changing the polarization of the polarized component) composed with an ope\-ra\-tor of a distinct kind.
For example let us consider a partial depolarizer $D_{\rm p}$ (in the meaning of the previous section) and a birefringent of axis ${\q e}_m$ and birefringence $\varphi $.
The whole device, denoted $D'_{\rm p}$, is such that, for $X=X_0(1+\I\rho\,{\q e}_n)$
\begin{eqnarray}
  D'_{\rm p}(X)&=&
  \exp\left({\q e}_m{\varphi \over 2}\right)\,D_{\rm p}(X)\,
  \exp\left(-{\q e}_m{\varphi \over 2}\right) \nonumber \\
  &=&\exp\left({\q e}_m{\varphi \over 2}\right)D_{\rm p}[X_0(1+\I\rho\,{\q e}_n)]
  \exp\left(-{\q e}_m{\varphi \over 2}\right)\nonumber \\
  &=& \exp\left({\q e}_m{\varphi \over 2}\right)[X_0(1+\I\gamma\rho\,{\q e}_n)]
  \exp\left(-{\q e}_m{\varphi \over 2}\right)\nonumber \\
  &=& X_0(1+\gamma \rho {\q e}_p)\,,
\end{eqnarray}
where
\begin{equation}
  {\q e}_p= \exp\left({\q e}_m{\varphi \over 2}\right)\,{\q e}_n\, \exp\left(-{\q e}_m{\varphi \over 2}\right)\,,\end{equation}
is deduced from ${\q e}_n$ in the rotation of angle $\varphi$ around ${\q e}_m$.

It shoud be clear that the depolarizer $D_{\rm p}$ and the birefringent device commute, that is,
  \begin{equation}
   \exp\left({\q e}_m{\varphi \over 2}\right)D_{\rm p}(X)
  \exp\left(-{\q e}_m{\varphi \over 2}\right)= D_{\rm p} \left[\exp\left({\q e}_m{\varphi \over 2}\right)\,X\,
    \exp\left(-{\q e}_m{\varphi \over 2}\right)\right]\,.\nonumber \\
  \end{equation}

  The depolarizing effect is due to $D_{\rm p}$ only, and changing the direction of polarization  (from ${\q e}_n$ to ${\q e}_p$) is due to the birefringent device only.
 
\section{Link with Azzam's orthogonalization problem}

We recall the Azzam's problem \cite{Azz}: given an arbitrary completely-polarized wave, find a device that transforms it into an orthogonally-polarized wave. A half-wave plate with appropriate orientation provides the
simplest solution \cite{Sol}. However, we may extend the issue to partially polarized lightwaves and look for a device that transforms the polarized component of a partially polarized state into its orthogonal state (with the same power). The solution is plain, because if a half-wave plate transforms the completely-polarized state $X_0(1+\I\,{\q e}_n)$ into the orthogonal state $X_0(1-\I\,{\q e}_n)$, it also transforms  $\rho X_0(1+\I\,{\q e}_n)$ into $\rho X_0(1-\I\,{\q e}_n)$ (and it preserves the unpolarized component $(1-\rho)X_0$). Every solution of the Azzam's issue also works for a partially polarized wave.

We now point out that a  given half-wave plate is a solution to the Azzam's issue only for incident polarization-states  whose representative points lie on the great circle whose plane is orthogonal to the plate axis (on the Poincar\'e sphere). In the following, we provide a ``weak'' solution to the extended Azzam's issue  that
holds for every incident polarization-state, but without preserving the power of the polarized component.

If $X=X_0(1+\I\rho \,{\q e}_n)$, then $-{\q e}_1X{\q e}_1-{\q e}_2X{\q e}_2-{\q e}_3X{\q e}_3=X_0(3-\I\rho\,{\q e}_n)$, whose polarized component is $\rho X_0(1-\I\,{\q e}_n)$, orthogonal to the polarized component  of $X$ 
(see Remark~1). But the power of $X_0(3-\I\rho\,{\q e}_n)$ is three times the power of $X$. Unless we use light amplifiers, we have to limit ourselves to passive devices.

A passive device performing the
mapping 
\begin{equation}
A\; : \; X\longmapsto A(X)=-{1\over 3}({\q e}_1\,X\,{\q e}_1+{\q e}_2\,X\,{\q e}_2+{\q e}_3\,X\,{\q e}_3)\,,\end{equation}
thus provides a  weak solution to the extended Azzam's problem, that holds for every incident partially-polarized state $X=X_0(1+\I\rho\,{\q e}_n)$: the polarized component of state $A(X)=X_0(3-\I\rho\, {\q e}_n)/3$
is $\rho X_0(1-\I\,{\q e}_n)/3$ and is orthogonal to $\rho X_0(1+\I\,{\q e}_n)$, the polarized component of $X$.
There is a price to pay for that result: the unpolarized component of $A(X)$ is $[1-(\rho /3)] X_0$ and is greater than  $(1-\rho )X_0$, the unpolarized component of $X$ (because $\rho > 0$). Part of the incident polarized power has been transformed into unpolarized power.

An actual device can be
deduced from the setup of Fig.\ 2: we only have to suppress BS~0 and  beam 0.


\begin{thebibliography}{10}


\bibitem{PPF1} P. Pellat-Finet, Repr\'esentation des \'etats et des op\'erateurs de
          polarisation de la lumi\`ere par des quaternions, {\em Optica Acta} {\bf 31} (1984) 415--434.

\bibitem{PPF2} \noindent{P. Pellat-Finet}, An introduction to a vectorial calculus for
polarization optics,  {\em Optik} {\bf 84} (1990) 169--175. 

\bibitem{PPF3} {P. Pellat-Finet, M. Bausset}, What  is common to both polarization
  optics and relativistic kinematics? {\em Optik}  {\bf 90} (1992) 101--106.

\bibitem{PPF4} P. Pellat-Finet, Geometrical approach to  polarization
  optics. {\sc I}-- Geometrical structure of polarized light,
  {\em Optik} {\bf 87} (1991) 27--33.

\bibitem{PPF5} P. Pellat-Finet, Geometrical approach to  polarization
  optics. {\sc II}-- Quaternionic representation  of polarized light,
  {\em Optik} {\bf 87} (1991) 68--76.

 \bibitem{Bor} {M.\ Born, E.\ Wolf}, {\em Principles of Optics}, 7th  edn,
        Cambridge University Press, Cambridge, 1999.

\bibitem{Ram} {G.\ N. Ramachandran, S. Ramaseshan}, {\em Crystal Optics}, in
  Handbuch der Physik {\sc XXV}/1, S. Fl\"ugge editor, Springer Verlag,
  Berlin, 1961.

\bibitem{PPF0} P. Pellat-Finet, {\sl Optique de Fourier, th\'eorie m\'etaxiale et fractionnaire}, Springer, Paris, 2009.
  

  \bibitem{Syn} J.\ L.\ Synge, Quaternions, Lorentz transformations, and the Conway-Dirac-Eddington matrices, Dublin Institute for Advanced Studies, Serie {\bf A 21} (1972) 1--67.

\bibitem{Azz}  { R. M. A. Azzam}, Polarization orthogonalization properties of optical
   systems, {\em Appl. Phys.} {\bf 13} (1977) 281--285.

\bibitem{Sol} The Azzam's issue is almost trivial when solved with the Poincar\'e sphere, if we look for a birefringent device as a solution. Let $P$ denote the point on the sphere that represents the incident polarization state and let $P_\perp$ represent the orthogonal  state. Points $P$ and $P_\perp$ are opposite on the sphere. Every (elliptical) birefringent transforms $P$ into a point $P'$ that is on a circle containing $P$ and whose plane is orthogonal to  the birefringent axis. For obtaining $P'$ at $P_\perp$, the circle is necessarily a great circle and the rotation angle is $\pi$. The birefringent is a half-wave plate (an elliptical one in general). Let ${\cal C}$ be a great circle passing by $P$ and $P_\perp$. The end points of the diameter of the sphere, orthogonal to the plane containing ${\cal C}$, provide the eigenstates of the half-wave plate. If $P$ and $P_\perp$ are not the North and South poles of the Poincar\'e sphere, then only one circle ${\cal C}$ (passing by $P$ and $P_\perp$) is in a plane orthogonal to the equatorial plane, and there is only one solution with a rectilinear half-wave plate (having rectilinear eigenvibrations). If $P$ and $P_\perp$ are the North and South poles, then every rectilinear half-wave plate is a solution. We eventually remark that a half-wave plate is solution of the Azzam's issue for all polarization states whose representative points are on the great circle whose plane is orthogonal to the plate axis.

\end{thebibliography}
\end{document}